\documentclass{PoS}

\usepackage{amsmath,amssymb,bm}
\usepackage{latexsym}
\newcommand{\ket}{\,\rangle}
\newcommand{\bra}{\langle \,}

\newcommand{\U}{\text{U}}
\newcommand{\mathd}{\mathrm{d}}

\newcommand{\mA}{\mathcal{A}}
\newcommand\tr{{\rm tr }}

\newcommand{\mF}{\mathcal{F}}

\title{Holography, chiral Lagrangian and form factor relations}

\ShortTitle{Holography, chiral Lagrangian and form factor relations}

\author{\speaker{Fen Zuo}
         \\
        Istituto Nazionale di Fisica Nucleare, Sezione di Bari\\
        E-mail: \email{fen.zuo@ba.infn.it}}

\abstract{We perform a detailed study of mesonic properties in a class of holographic models of QCD, which is described by the Yang-Mills plus Chern-Simons action. By decomposing the 5 dimensional gauge field into resonances and integrating out the massive ones, we reproduce the Chiral Perturbative Theory Lagrangian up to ${\cal O}(p^6)$ and obtain all the relevant low energy constants~(LECs). The numerical predictions of the LECs show minor model dependence, and agree reasonably with the determinations from other approaches. Interestingly, various model-independent relations appear among them. Some of these relations are found to be the large-distance limits of universal relations between form factors of the anomalous and even-parity sectors of QCD.}

\FullConference{Xth Quark Confinement and the Hadron Spectrum,\\
		October 8-12, 2012\\
		TUM Campus Garching, Munich, Germany}

\renewcommand{\logo}
{\parbox{\textwidth}{\begin{flushright} \vspace*{2cm}
 BARI-TH-2013-667
\end{flushright}
\includegraphics{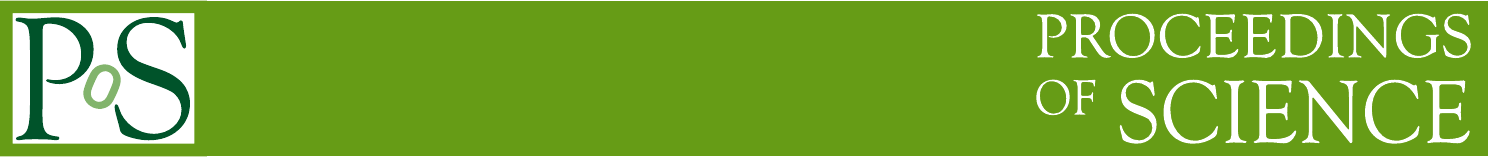}}
}
\begin{document}

\section{Introduction}
In recent years the holographic approach towards QCD, based on the gauge/string duality~\cite{Maldacena:1997re,Gubser:1998bc,Witten:1998qj}, has provided us with many novel ideas and promising results. Of particular interest is a class of models described by the Yang-Mills~(YM) and Chern-Simons~(CS) action~\cite{Son:2003et}, which comes out as a large-$N_c$ generalization of the traditional hidden local symmetry model. In this framework a multiplet of Goldstone bosons is built in at the beginning, and chiral symmetry is realized in the non-linear pattern. Later a concrete model in this class was constructed by embedding $D8$ and ${\bar D}8$ flavor branes into the  background of $D4$ branes~\cite{Sakai:2004cn}. Chiral symmetry breaking is implemented geometrically in the configuration with $D8$ and ${\bar D}8$ branes connected smoothly in the infrared.

Since chiral symmetry breaking is quite naturally accommodated in this framework, the low energy properties
 are nicely reproduced. For example, truncated to the pion sector one reproduces the Chiral Perturbation Theory~($\chi PT$) Lagrangian up to ${\cal O}(p^4)$~\cite{Sakai:2004cn,Hirn:2005nr,Sakai:2005yt}. In particular, the Chern-Simons part immediately gives rise to the Wess-Zumino-Witten term~\cite{Sakai:2005yt}.
  In the even parity sector, the relevant low energy constants~(LECs) are quite accurately determined, showing little model dependence~\cite{Sakai:2004cn,Hirn:2005nr,Sakai:2005yt}. Recently, the anomaly structure in these models was further studied~\cite{Son:2010vc}, and a universal relation for the transverse part of triangle anomalies was found. Studies of such a relation were carried out in \cite{Knecht:2011wh,Colangelo:2011xk,Gorsky:2012ui}. These progress stimulated us to perform a systematic investigation within this class of models~\cite{Colangelo:2012ip}. Here I summarize all the results along these lines.

\section{Review of the holographic framework}
\subsection{The 5D picture and correlations functions}
The class of models we focus on are described by the action~\cite{Son:2003et,Sakai:2004cn}
\begin{eqnarray}
  S &=& S_{\rm YM}+S_{\rm CS}
  \label{eq.5Daction}
  \\
  \label{eq:YM}
  S_{\rm YM} &=& -\int\! d^5x \tr \left[-f^2(z){\cal F}_{z\mu}^2
  + \frac{1}{2g^2(z)}{\cal F}_{\mu\nu}^2 \right], \\
  \label{eq:CS}
  S_{\rm CS} &=& -\kappa \int\! \tr
  \left[{\cal AF}^2+\frac{i}{2}{\cal A}^3{\cal F}-\frac{1}{10}{\cal A}^5
\right].
\end{eqnarray}
Here ${\cal A}(x,z)={\cal A}_M dx^M$ is the 5D $\U(N_f)$ gauge field and
${\cal F}=d{\cal A}-i{\cal A} \wedge {\cal A}$ is the field strength.
They are decomposed as ${\cal A}= {\cal A}^a t^a$ and
${\cal F}= {\cal F}^a t^a$, with the normalization of the generators Tr$\{t^at^b\}=\delta^{ab}/2$.
The coefficient $\kappa=N_C/(24\pi^2)$, with $N_C$ the number of colors.
The functions $f^2(z)$ and $g^2(z)$ are invariant under reflection $z\to -z$
so that parity can be properly defined in the model. The fifth coordinate $z$ runs from $-z_0$ to $z_0$ with $z_0>0$. $z_0$ can be finite or infinite depending on the backgrounds.

To calculate the correlation functions, one needs to know the bulk-to-boundary propagators, which describe the response of the system to external sources. In the present case the most interesting ones are those for the transverse part of the gauge field, which can be further decomposed into the vector part and the axial part. To obtain them first one derives the equation of motion from the action, which after the 4D Fourier transformation becomes
\begin{equation}
g^2(z)\, \partial_z[f^2(z)\partial_z {\cal A}_\mu(q ,z)]
\, =\, -q^2 {\cal A}_\mu(q ,z) \, .\label{eq.5D-EoM}
\end{equation}
Requiring the boundary conditions $V(Q,\pm z_0)=1$ and
$A(Q,\pm z_0)=\pm 1$~($Q^2\equiv -q^2$), one then gets the explicit expressions of them.

The original 5D gauge potential can be decomposed as an evolution of the bulk-to-boundary propagator with the external source. Various correlations functions are then obtained by taking the functional derivative in the 5D action,  with respect to the corresponding sources. For example, from the Yang-Mills term one finds the two-point vector and axial current correlators
\begin{eqnarray}
\Pi_V(Q^2)&=&\frac{1}{Q^2}f^2(z)V(Q,z)\partial_z V(Q,z)|_{z=-z_0}^{z=+z_0}\nonumber\\
\Pi_A(Q^2)&=&\frac{1}{Q^2}f^2(z)A(Q,z)\partial_z A(Q,z)|_{z=-z_0}^{z=+z_0}.
\end{eqnarray}
From the Chern-Simons term one obtains the longitudinal and transverse part of the triangle anomaly~(in the kinetic limit where one vector field is soft)~\cite{Son:2010vc}
\begin{equation}
w_L(Q^2)=\frac{2N_c}{Q^2},\quad w_T(Q^2)=\frac{N_C}{Q^2}\int_{-z_0}^{z_0}\mathd z ~A(Q,z)\partial_z V(Q,z).
\end{equation}
Taking into account that $V(Q,z)$ and $A(Q,z)$ are the two independent solutions of eq.~(\ref{eq.5D-EoM}) with different boundary conditions, one obtains a novel model-independent relation~\cite{Son:2010vc}
\begin{equation}
w_T(Q^2)=\frac{N_C}{Q^2} + \frac{N_C}{f_\pi^2}\,
[ \Pi_V(Q^2)  -  \Pi_A(Q^2) ]\, .\label{eq.SY}
\end{equation}

\subsection{The 4D picture and $\chi PT$ Lagrangian}
The equation of motion (\ref{eq.5D-EoM}) also has normalizable solutions $\psi_n$ for discrete values $q^2=m_n^2$, which correspond to vector and axial meson states depending on the property under the transformation $z\to -z$. The Goldstone bosons are contained in the gauge component ${\cal A}_z$
and can be parameterized through the chiral field $U$ as
\begin{equation}
U(x^\mu)=\mbox{P} \exp\left\{i\int^{+z_0}_{-z_0} {\cal A}_z(x^\mu,z') dz'\right\}.\label{eq.U}
\end{equation}
One can further introduce the external sources as the boundary values of the gauge potential ${\cal A}_\mu$, and treat them as if they are dynamical~\cite{Son:2003et,Sakai:2005yt}. The propagation of these source fields in the fifth dimension is then controlled by the zero mode solutions of (\ref{eq.5D-EoM}), namely $1=V(0,z)$ and $\psi_0(z)=A(0,z)$. With all these ingredients included, one finds the on-shell decomposition of the gauge potential ${\cal A}_\mu$,
\begin{equation}
{\cal A}_\mu(x,z)=\ell_\mu(x) \psi_-(z)+r_\mu(x) \psi_+(z)
+\sum_{n=1}^\infty v_\mu^n(x)\psi_{2n-1}(z)
+\sum_{n=1}^\infty a_\mu^n(x)\psi_{2n}(z),\,
\label{eq.Amu-decomposition1}
\end{equation}
where $\ell_\mu(x)$ and $r_\mu(x)$ are the source fields at the left and right boundaries, and $\psi_\pm(z)=\frac{1}{2}(1\pm\psi_0(z))$. Moreover, one can make the formulas more compact employing the ${\cal A}_z=0$ gauge, in which eq.~(\ref{eq.U}) and eq.~(\ref{eq.Amu-decomposition1}) are combined into a single expression
\begin{equation}
{\cal A}_\mu(x,z)=i\Gamma_\mu(x)+\frac{u_\mu(x)}{2}\psi_0(z)
+\sum_{n=1}^\infty v_\mu^n(x)\psi_{2n-1}(z)
+\sum_{n=1}^\infty a_\mu^n(x)\psi_{2n}(z)\, ,
\label{eq.Amu-decomposition2}
\end{equation}
with the commonly used tensors $\Gamma_\mu(x)$ and $u_\mu(x)$ in $\chi PT$.

We can rewrite the 5D action in a 4D form. As a first step, one sets all the massive fields to be
zero and focuses on the pion sector. Substituting the $\mA_\mu$ decomposition~(\ref{eq.Amu-decomposition2})
in the YM action, one gets the even-parity $\chi PT$ Lagrangian up to ${\cal O}(p^4)$. The LECs in the Lagrangian are given by the 5D integrals:
\begin{eqnarray}
f_\pi^2&=&4\left(\int_{-z_0}^{z_0}\frac{\mathd z}{f^2(z)}\right)^{-1} \, ,\qquad
L_1=\frac{1}{2} L_2=-\frac{1}{6}L_3=\frac{1}{32}\int_{-z_0}^{z_0} \frac{(1-\psi_0^2)^2}{g^2(z)}  \, \mathd z \, ,
\nonumber\\
L_9&=&-L_{10}=\frac{1}{4}\int_{-z_0}^{z_0} \frac{1-\psi_0^2}{g^2(z)}  \, \mathd z \, ,\qquad\qquad H_1=-\frac{1}{8}\int_{-z_0}^{z_0} \frac{1+\psi_0^2}{g^2(z)}   \, \mathd z \, .
\end{eqnarray}
Notice that at this order, there are already model-independent relations among the LECs. Furthermore, the substitution of the $\mA_\mu$ decomposition~(\ref{eq.Amu-decomposition2})
in the Chern-Simons action reproduces the gauged Wess-Zumino-Witten term~\cite{Sakai:2004cn,Sakai:2005yt}. 

\section{$\chi PT$ Lagrangian at ${\cal O} (p^6)$}
In ref.~\cite{Colangelo:2012ip} we explore further the predictions along these two lines within this class of models. First, we notice that the results from the two formalisms are not independent. For example, one can  calculate the quantities on both sides of the relation (\ref{eq.SY}) using the resonance decomposition (\ref{eq.Amu-decomposition2}). The relation turns into an infinite number of matching conditions among the resonance parameters. In particular, taking the $Q^2\to 0$ limit, the relation becomes~\cite{Knecht:2011wh}
\begin{equation}
C_{22}^W= - \frac{N_C}{32\pi^2 f_\pi^2} L_{10},\label{eq.Cw22}
\end{equation}
where $C_{22}^W$ is an ${\cal O} (p^6)$ LEC in the odd-parity sector~\cite{Bijnens:2001bb}. While we have shown before that $\chi PT$ Lagrangian up to ${\cal O} (p^4)$ can be reproduced directly, no derivation for the higher order terms has been done. Phenomenologically, although the full set of independent operators have been constructed about ten years ago~\cite{Bijnens:1999hw,Ebertshauser:2001nj,Bijnens:2001bb}, accurate determination of the coefficients are still difficult.

 Based on the derivation in the previous section, we know that these higher order terms can only come from the resonance exchanging diagrams. To reproduce the ${\cal O} (p^6)$ operators, only the one-resonance interactions are needed since terms with more resonance fields only contribute to operators of even higher order.
Substituting the decomposition (\ref{eq.Amu-decomposition2}) into the 5D action, one can extract the parts
\begin{eqnarray}
S_{\rm{YM}}\bigg|_{{\rm Kin.}}&=&\sum_n \int \mathd ^4x
\bra    -\frac{1}{2} (\nabla_\mu v_\nu^n-\nabla_\nu v_\mu^n)^2
+m_{v^n}^2 {v_\mu^n}^2
-\frac{1}{2}(\nabla_\mu a_\nu^n-\nabla_\nu a_\mu^n)^2
+m_{a^n}^2 {a_\mu^n}^2\ket\, ,
\label{eq.S-kin} \\
\nonumber\\
S_{\rm{YM}}\bigg|_{{\rm 1-res.}}&=&\sum_n \int \mathd ^4x
\, \bigg\{\,
-\,  \bra \frac{f^{\mu\nu}_+}{2}
\bigg[    (\nabla_\mu v_\nu^n-\nabla_\nu v_\mu^n)a_{Vv^n}-\frac{i}{2}([u_\mu,a_\nu^n]
\, -\, [u_\nu,a_\mu^n])a_{Aa^n}  \bigg] \ket
\label{eq.S-1res-even} \\
&&
\qquad  -\, \bra
\frac{i}{4}[u^\mu,u^\nu]
\bigg[    (\nabla_\mu v_\nu^n-\nabla_\nu v_\mu^n)b_{v^n\pi\pi}
\, -\, \frac{i}{2}([u_\mu,a_\nu^n]-[u_\nu,a_\mu^n])b_{a^n\pi^3}]\ket
\nonumber\\
&&  \qquad  +\,   \bra  \frac{f^{\mu\nu}_-}{2}
\bigg[  (\nabla_\mu a_\nu^n-\nabla_\nu a_\mu^n)a_{Aa^n}
\, -\, \frac{i}{2}([u_\mu,v_\nu^n]-[u_\nu,v_\mu^n])(a_{Vv^n}-b_{v^n\pi\pi})\bigg]\ket
\,\bigg\}\, ,
\nonumber\\
S_{\rm{CS}}\bigg|_{{\rm 1-res.}}&=&
\sum_n \int \mathd ^4x \,
\bigg\{
-\frac{N_C}{32\pi^2}\, c_{v^n}\epsilon^{\mu\nu\alpha\beta}
\bra u_\mu\{v_\nu^n,f_{+\alpha\beta}\}\ket
+ \frac{N_C}{64\pi^2}\, c_{a^n}\epsilon^{\mu\nu\alpha\beta}
\bra u_\mu\{a_\nu^n,f_{-\alpha\beta}\}\ket
\nonumber \\
&&\qquad \qquad \qquad +\frac{i N_C}{16\pi^2}\, (c_{v^n}-d_{v^n})\epsilon^{\mu\nu\alpha\beta}
\bra v_\mu^n u_\nu u_\alpha u_\beta\ket \,\bigg\} \, .
\label{eq.S-1res-odd}
\end{eqnarray}
Here some notations and terms in $\chi PT$ have been used, and all the couplings are given by the integral of the corresponding wave functions $\psi_0$ and $\psi_n$ over $z$. In particular, one finds that the couplings from the YM part and CS part are related due to the equation of motion
\begin{eqnarray}
c_{v^n}&=&\frac{m_{v^n}^2}{2f_\pi^2 }b_{v^n\pi\pi}\, , \qquad\qquad
c_{a^n}=\frac{m_{a^n}^2}{3f_\pi^2 }b_{a^n\pi^3}\, .
\label{eq.cvca}
\end{eqnarray}
It turns out that these relations are essential for the validation of the form factor relations we show in the next section.

Since we are working at $N_C\to \infty$, in the above action we have the interactions of infinite number of resonances. One would like to know how it can be approximated with only a few or even one resonance. From a simple model with the "cosh" metric function~\cite{Son:2003et}, one can see explicitly how the approximation works. The model is specified by
\begin{equation}
f^2(z)=\Lambda^2 \cosh^2(z) /g_5^2, ~~g^2(z)=g_5^2,~~ z_0=\infty,
\end{equation}
and all the independent couplings in this model read
\begin{eqnarray}
a_{Vv^n}&=&a_{2n-1},~~~~a_{Aa^n}=a_{2n},~~~a_n=\frac{1}{g_5}\sqrt{\frac{2(2n+1)}{n(n+1)}},\nonumber\\
c_{v^n}&=&\frac{g_5}{\sqrt{3}}\delta_{n,1},~~~c_{a^n}=\frac{2g_5}{\sqrt{15}}\delta_{n,1},\nonumber\\
d_{v^n}&=&\frac{\sqrt{3}g_5}{15}\delta_{n,1}+\frac{2\sqrt{42}g_5}{105}\delta_{n,2},
\end{eqnarray}
One sees that the approximation with the first one or two resonances is accurate for most couplings, the only exception being the couplings to the external sources, $a_{Vv^n}$ and $a_{Aa^n}$. This is because we need an infinite number of resonances to reproduce the logarithmic behavior of the corresponding correlators at large $Q^2$~\cite{Shifman:2005zn}.

The ${\cal O} (p^6)$ $\chi PT$ operators result from integrating out the resonances from the above Lagrangian, with the coefficients given by the summation of the resonance couplings. For example, in the odd-parity sector, one finds
\begin{equation}
C_{22}^W=\frac{N_C}{64\pi^2}\sum_{n=1}^\infty \frac{a_{Vv^n}c_{v^n}}{m_{v^n}^2}.
\end{equation}
Using the coupling relation (\ref{eq.cvca}) and the completeness condition for the solutions $\psi_n$, this can be further simplified and finally becomes (\ref{eq.Cw22}). In a similar way, all the other LECs in the odd sector can be expressed through the ${\cal O} (p^4)$ couplings $L_1,...,L_{10}$, together with an additional parameter $Z$
\begin{eqnarray}
Z &\equiv& \int_{-z_0}^{+z_0} \frac{\psi_0^2(1-\psi_0^2)^2}{4g^2(z)}  \, \mathd z \,.
\label{eq.Z-def}
\end{eqnarray}
For example, the operator $O_{23}^W$~\cite{Bijnens:2001bb}, which is obtained by replacing the vector source in $O_{22}^W$ with the axial one, has the coefficient
\begin{equation}
C_{23}^W=\frac{N_C}{128\pi^2}\sum_{n=1}^\infty \frac{a_{Aa^n}c_{a^n}}{m_{a^n}^2}=\frac{N_C}{96\pi^2f_\pi^2}(L_9-8L_1).
\label{eq:Cw23}
\end{equation}
In the even-parity sector, one obtains similar resonance expressions for all the LECs. In particular, when the contributions are from two odd vertexes, the expressions can be further simplified as for $C_{22}^W$. Finally these odd-odd terms are completely determined by $f_\pi$ and $N_C$, e.g,
\begin{equation}
C_{52}=-\frac{1}{8}\sum_{n=1}^\infty\frac{a_{Vv^n}b_{v^n\pi\pi}}{m_{v^n}^2}-\frac{N_C^2}{1920\pi^4f_\pi^2}.
\end{equation}
Many relations among the even couplings at this order exist, extending the previously found relations at ${\cal O} (p^4)$. Some of the most interesting ones are
\begin{eqnarray}
&&3 C_3 + C_4  = C_1 + 4 C_3\, ,\label{eq.C1-4}
\\
&&2 C_{78} - 4 C_{87} + C_{88}  = 0\, ,\label{eq.C78-88}\\
&&a^{00}_2=N_C^2\,,\,~~~~b^{00}=\frac{N_C^2}{6}\,,\label{eq.ab00}\\
&&a^{+-}_2=0\,,\,~~~~~~~~~~b^{+-}=0.\label{eq.ab+-}
\end{eqnarray}
Here $a^{00}_2,~b^{00}$ and $a^{+-}_2,~b^{+-}$ are combinations of LECs relevant for the $\gamma\gamma\to \pi^0\pi^0$ and $\gamma\gamma\to \pi^+\pi^-$ processes, respectively.
Numerically, these relations are satisfied reasonable well~\cite{Colangelo:2012ip}, see e.g., Tab.~\ref{tab.ab}.

\begin{table}[t!]
\centering
\begin{tabular}{|c|c|c|c|c|}
\hline
   ~~~~~~~       &   Holo. &  DSE   &  Reso.  Lagr.  & ENJL   \\\hline
$a^{00}_2$       &    $N_C^2$    &  $3.79$ & $13\pm 3.3$& $14.0$  \\
$b^{00}$         &    $N_C^2/6$  &  $1.66$ & $3\pm1$   &$1.66$   \\\hline
 $a^{+-}_2$   &   $0$     &  $-0.98$  & $0.75\pm 0.65$  & $6.7$ \\
  $b^{+-}$    &   $0$     & $-0.23$  & $0.45\pm 0.15$ & $0.38$ \\\hline
\end{tabular}
\caption{Holographic predictions of the parameters $a_2$ and $b$ relevant for the $\gamma\gamma\to \pi\pi$ scattering processes, in comparison with the results from the Dyson-Schwinger Equation approach, the resonance Lagrangian, and the extended Nambu-Jona-Lasinio model. More details can be found in ref.~\cite{Colangelo:2012ip}.}\label{tab.ab}
 \end{table}

\section{Form factor relations}
It turns out that some of these relations between LECs can be generalized to relations between scattering amplitudes, correlation functions or form factors. As discussed in ref.~\cite{Hirn:2005nr}, the large-$N_C$ assumption and the fact that only vector/axial resonances are included give strict constraints on the $\pi\pi$ scattering amplitude. As a result, one finds the relation $L_1=\frac{1}{2} L_2=-\frac{1}{6}L_3$ at ${\cal O} (p^4)$, and further the relation (\ref{eq.C1-4}) among $C_1,...,C_4$ at higher order. Similar reasoning for the $\gamma\gamma\to \pi\pi$ processes gives rise to relations (\ref{eq.ab00}) and (\ref{eq.ab+-})~\cite{Colangelo:2012ip}. The relation $L_9=-L_{10}$ is found to be related to the vanishing of the axial form factor in the $\pi\to l\nu\gamma$ decay~\cite{Hirn:2005nr}, which at higher order results in (\ref{eq.C78-88}). As for the relation between $C_{22}^W$ and $L_{10}$ (\ref{eq.Cw22}), we have already shown that it results from the relation (\ref{eq.SY}) between different correlation functions.

\begin{figure}[ht]
\centering
	\includegraphics[width=0.50\textwidth]{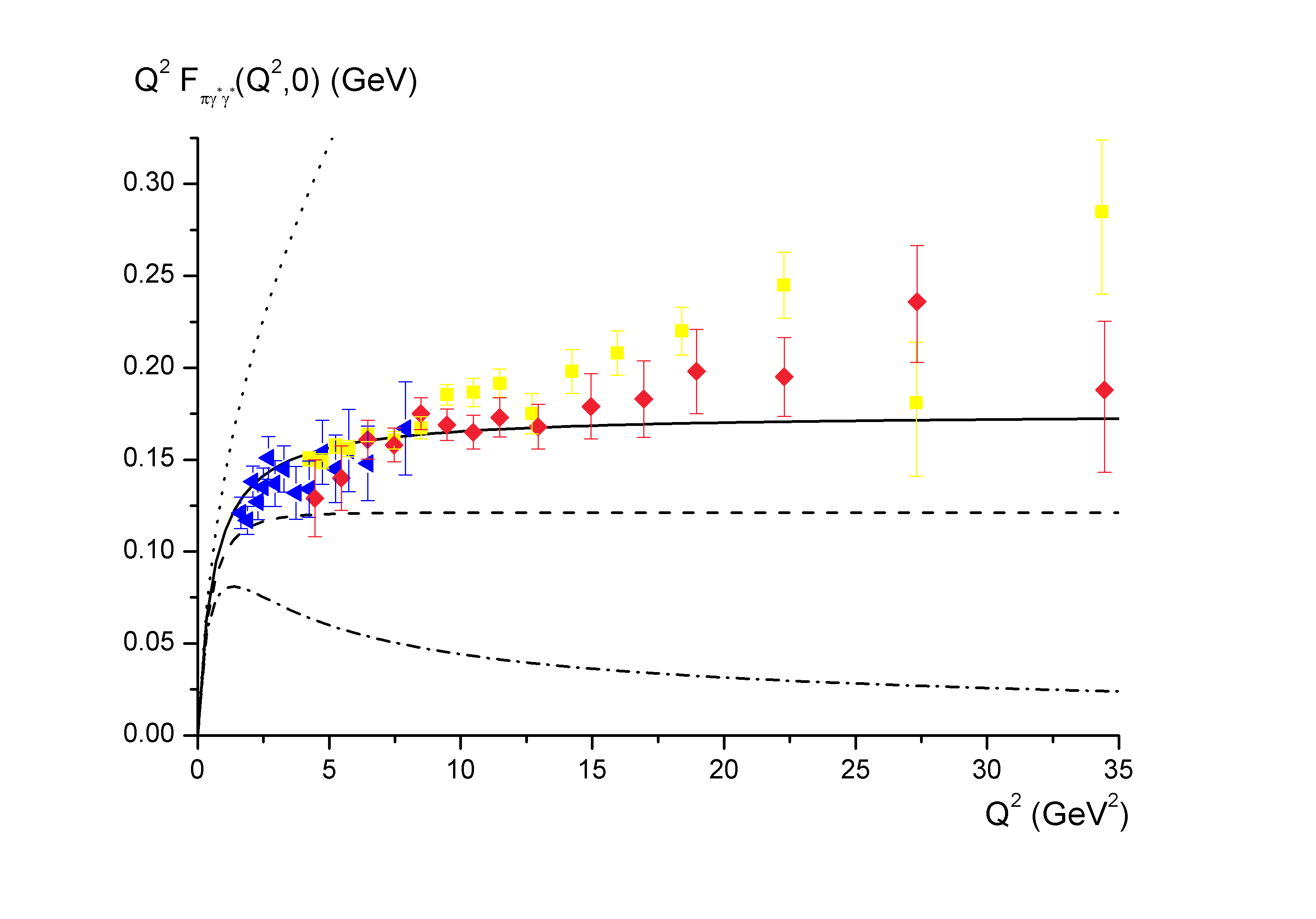}\includegraphics[width=0.5\textwidth]{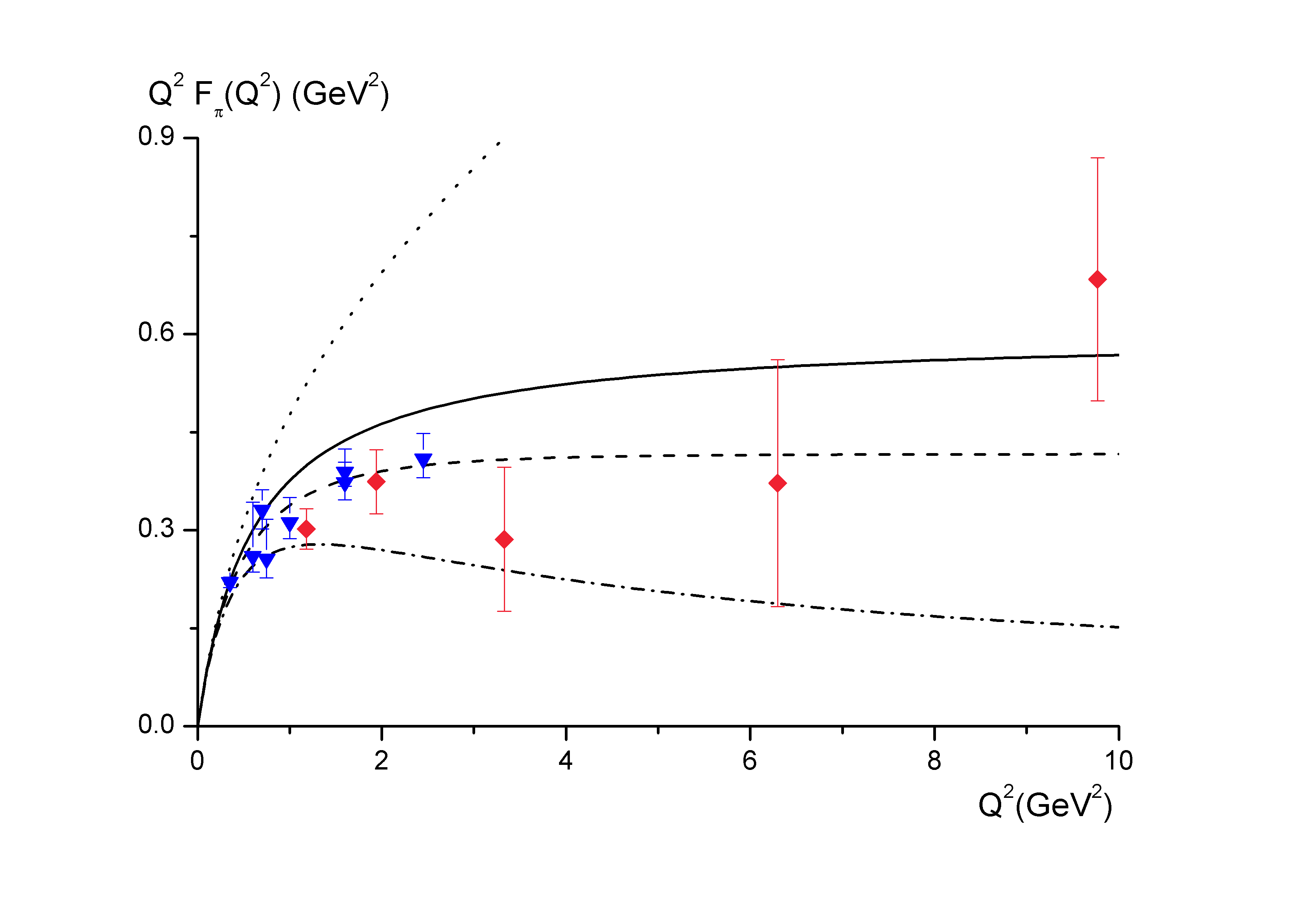}
\caption{\it Anomalous $\pi\gamma^*\gamma^*$ form factor and the electromagnetic pion form factor calculated  from the flat, ``cosh'', hard wall and Sakai-Sugimoto models,
denoted by the dotted, solid, dashed and dash-dotted lines, respectively. For the details of different models and the experimental data, please see ref.~\cite{Colangelo:2012ip}.}\label{fig:FF}
\end{figure}

 What about the remaining relations of the other LECs? What are the underlying reasons for them? Or could they also be generalized to relations valid in the whole momentum region? From to the above mentioned relations, an immediate observation is that $C_{22}^W$ and $L_9$ are also related to each other. Thus a reasonable guess will be that there could be some relation between the anomalous $\pi\gamma^*\gamma^*$ form factor and the electromagnetic pion form factor, which at low energy are related to $C_{22}^W$ and $L_9$ respectively. These form factors can be calculated either in the 5D formalism, or using the 4D resonance decomposition. In the 5D picture, the results are more compact and read
\begin{eqnarray}
\mF_{\pi\gamma^*\gamma^*}(Q_1^2,Q_2^2)&=&\frac{N_C}{24\pi^2f_\pi}\int_{-z_0}^{z_0} V(Q_1,z) V(Q_2,z)\partial_z\psi_0(z)\mathd z,\nonumber\\
\mF_\pi(Q^2)&=&\frac{1}{f_\pi^2}\int_{-z_0}^{z_0} f^2(z) V(Q,z) [\partial_z\psi_0(z)]^2  \mathd z.
\end{eqnarray}
Employing the equation of motion for $\psi_0$ one finds
\begin{equation}
\mF_{\gamma^*\gamma^*\pi}(Q^2,0)=\frac{N_C}{12\pi^2f_\pi}~ \mF_\pi(Q^2).\label{eq.VFFTFF}
\end{equation}
In Fig.~\ref{fig:FF} we show the results for the two form factors calculated from different models, and compare them with the experimental data. One finds that the ``cosh'' and hard wall models, in which the backgrounds are asymptotic anti-de Sitter, are able to fit the data in the large $Q^2$ region. However, more accurate data for $\mF_\pi(Q^2)$ are needed to check if the relation (\ref{eq.VFFTFF}) could be valid or not. Replacing the photon by the axial source, one gets a relation as (\ref{eq.VFFTFF}), which at low energy reduces to (\ref{eq:Cw23}).

\section{Summary}
I reviewed our results obtained in ref.~\cite{Colangelo:2012ip} for the interactions among the pions, vector/axial mesons and external gauge sources within a class of holographic models. The Lagrangian with one resonance field is derived and from this we obtain all the ${\cal O} (p^6)$ LECs. Various model-independent relations among these LECs are found. Inspired by these results, we found some interesting relations between the form factors with different intrinsic parity. Further study along these lines is still in progress.

     \acknowledgments
       \small
        I thank Pietro Colangelo and Juan Jose Sanz-Cillero for the collaboration. This work was supported by the Italian MIUR PRIN 2009 and the National Natural Science Foundation of China under Grant No. 11135011.


\begin{thebibliography}{99}
\bibitem{Maldacena:1997re}
Juan~Martin Maldacena.
\newblock Adv.Theor.Math.Phys. \textbf{2} (1998): 231-252.
\bibitem{Gubser:1998bc}
S.S. Gubser, Igor~R. Klebanov, and Alexander~M. Polyakov.
\newblock Phys.Lett. \textbf{B428} (1998): 105-114.

\bibitem{Witten:1998qj}
Edward Witten.
\newblock Adv.Theor.Math.Phys. \textbf{2} (1998): 253-291.

\bibitem{Son:2003et}
D.~T. Son and M.~A. Stephanov.
\newblock Phys. Rev. \textbf{D69} (2004): 065020.

  \bibitem{Sakai:2004cn}
Tadakatsu Sakai and Shigeki Sugimoto.
\newblock Prog. Theor. Phys. \textbf{113} (2005): 843-882.

  \bibitem{Hirn:2005nr}
Johannes Hirn and Veronica Sanz.
\newblock JHEP \textbf{12} (2005): 030.

  \bibitem{Sakai:2005yt}
Tadakatsu Sakai and Shigeki Sugimoto.
\newblock Prog. Theor. Phys. \textbf{114} (2005): 1083-1118.


  \bibitem{Son:2010vc}

\newblock Dam~T. Son and Naoki Yamamoto (2010):
  [\href{http://arxiv.org/abs/1010.0718}{arXiv: 1010.0718}].


  \bibitem{Knecht:2011wh}
Marc Knecht, Santiago Peris, and Eduardo de~Rafael.
\newblock JHEP \textbf{1110} (2011): 048.


  \bibitem{Colangelo:2011xk}
P.~Colangelo, F.~De~Fazio, J.J. Sanz-Cillero, F.~Giannuzzi, and S.~Nicotri.
\newblock Phys.Rev. \textbf{D85} (2012): 035013.

  \bibitem{Gorsky:2012ui}
A.~Gorsky, P.N. Kopnin, A.~Krikun, and A.~Vainshtein.
\newblock Phys.Rev. \textbf{D85} (2012): 086006.

  \bibitem{Colangelo:2012ip}
P.~Colangelo, J.~J.~Sanz-Cillero and F.~Zuo.
\newblock JHEP \textbf{11} (2012): 012.

    \bibitem{Bijnens:2001bb}
J.~Bijnens, L.~Girlanda, and P.~Talavera.
\newblock Eur. Phys. J. \textbf{C23} (2002): 539-544.


  \bibitem{Bijnens:1999hw}
J.~Bijnens, G.~Colangelo, and G.~Ecker.
\newblock Annals Phys. \textbf{280} (2000): 100-139.



\bibitem{Ebertshauser:2001nj}
T.~Ebertshauser, H.W. Fearing, and S.~Scherer.
\newblock Phys.Rev. \textbf{D65} (2002): 054033.




\bibitem{Shifman:2005zn}
M.~Shifman.
  [\href{http://arxiv.org/abs/hep-ph/0507246}{arXiv: hep-ph/0507246}].














\end{thebibliography}
\end{document}